\documentclass[superscriptaddress,
showpacs,preprintnumbers,nofootinbib,amsmath,amssymb,aps,prd,11pt, groupedaddress]{revtex4-1}
\usepackage{amsmath}
\usepackage{graphics}
\usepackage{epstopdf}
\usepackage{subfigure}
\usepackage{braket}
\usepackage{bm}
\usepackage{dcolumn}
\usepackage{cancel}
\usepackage[dvipdfmx,final]{graphicx}
\input{colordvi.tex}
\usepackage{longtable}
\usepackage{tabularx}
\usepackage{cleveref}
\usepackage{mhchem}
\usepackage{mathtools}
\usepackage[utf8]{inputenc}
\usepackage{color}

\makeatletter
\renewcommand{\fnum@table}{\textbf{\tablename~\thetable}}
\renewcommand{\fnum@figure}{\textbf{\figurename~\thefigure}}
\makeatother

\newcommand {\be}{\begin{equation}}
\newcommand {\ee}{\end{equation}}
\newcommand {\ba}{\begin{eqnarray}}
\newcommand {\ea}{\end{eqnarray}}

\definecolor{darkgreen}{rgb}{0.0, 0.5, 0.0}  

\begin{document}


\vspace*{10mm}

\title{Uncovering Secret Neutrino Interactions at Tau Neutrino Experiments \vspace*{1.cm} }

\author{  Pouya Bakhti }
\author{  Meshkat Rajaee }
\author{  Seodong Shin }
\affiliation{Laboratory for Symmetry and Structure of the Universe, Department of Physics, Jeonbuk National University, Jeonju, Jeonbuk 54896, Korea}

\begin{abstract}
  \vspace*{.5cm}

We investigate the potential of future tau neutrino experiments
for identifying the $\nu_\tau$ appearance in probing secret neutrino interactions, which is very important in a variety of fields such as neutrino physics, dark matter physics, grand unified theories, astrophysics, and cosmology.
The reference experiments include the DUNE far detector utilizing the atmospheric data, which is for the first time in probing the secret interactions, 
the Forward Liquid Argon Experiment (FLArE100) detector at the Forward Physics Facility (FPF), and emulsion detector experiments such as SND@LHC, AdvSND, FASER$\nu$2, and SND@SHiP. 
For concreteness, we consider a reference scenario in which the hidden interactions among the neutrinos are mediated by a single light gauge boson $Z^\prime$ with a mass at most below the sub-GeV scale and an interaction strength $g_{\alpha \beta}$ between the active neutrinos $\nu_\alpha$ and $\nu_\beta$.
We confirm that these experiments have the capability to significantly enhance the current sensitivities on $g_{\alpha \beta }$ for $m_{Z^\prime} \lesssim 500$ MeV  
due to 
the production of high energy neutrinos and excellent ability to detect tau neutrinos. 
Our analysis highlights the crucial role of  {\it downward-going}
DUNE atmospheric data in the search for secret
neutrino interactions 
because of  the rejection of backgrounds dominated in the upward-going events. 
Specifically, 10 years of DUNE atmospheric data can 
provide the best sensitivities
on $g_{\alpha \beta}$ which is about two orders of magnitude improvement. 
In addition, 
the beam-based experiments such as FLArE100 and FASER$\nu$2 can improve the current constraint on $g_{e\tau}$ and $g_{\mu\tau}$ by more than an order of magnitude after the full running of the high luminosity LHC with the integrated luminosity of 3 ab$^{-1}$.
For $g_{e\mu}$ and $g_{ee}$ the SHiP experiment can 
play the most important role in the high energy region of $E>{\rm few}~100$ MeV.
Although our analysis is proceeded under our reference scenario of secret $Z'$, our analysis strategies
can be readily applicable to other types of secret interactions such as Majoron models.

\end{abstract}


\maketitle

\section{Introduction}

Among the particles in the Standard Model (SM), neutrinos are unique in the sense that they play the key roles not only in determining the weak interaction structures but also in guiding a new physics beyond the SM (BSM) due to their oscillation phenomena, not explained in the context of the SM.
Possible new interactions of neutrinos other than the weak interaction, therefore, can shed light on identifying the symmetrical structure of BSM.

One area of interest is the Secret Neutrino Interaction (SNI), which involves new 
 boson(s) mediating the interactions among the active and sterile neutrinos, or involving only the sterile neutrino sector. 
Secret neutrino interactions might 
arise in the BSM theories with the neutrino masses given from the breakdown of the 
global symmetries of the SM such as 
lepton number ($L$) or the difference between the baryon number and the lepton number ($B-L$) symmetries~\cite{Chikashige:1980ui,gelmini1981left,schechter1982neutrino,Choi:1991aa,Acker:1992eh,valle2015neutrinos}. 
Other possibilities include gauging an anomaly free global symmetry \cite{Araki:2012ip,Asai:2019ciz}, which is not technically ``secret", or introducing a new gauge symmetry completely blind to the SM particles~\cite{Farzan:2016wym}.

The theoretical scenarios providing SNI have been applied to explain the
neutrino oscillation anomalies~\cite{Asaadi:2017bhx,Smirnov:2021zgn,Dentler:2019dhz,Abdallah:2022grs,Muong-2:2006rrc,Dutta:2021cip}.
Interestingly, the SNIs have been also used to resolve various issues in cosmological and astrophysical observations.
The pseudo Nambu-Goldstone boson arising after the spontaneous break down of a global lepton number symmetry ($L$ or $B-L$) and the electroweak symmetry, so called Majoron, can be a dark matter (DM) candidate~\cite{Rothstein:1992rh}. 
The emission of Majorons or vector bosons can also contribute to the supernova cooling~\cite{Choi:1989hi, Akita:2022etk}.
Inclusion of SNI can make the thermal sterile neutrino DM scenario~\cite{Dodelson:1993je} compatible with the astrophysical observations~\cite{DeGouvea:2019wpf}. A new gauge boson mediating the SNI can be used to
resolve the small scale problems, albeit with strong cosmological constraints~\cite{vandenAarssen:2012vpm,Ahlgren:2013wba,Chu:2015ipa}, 
or alleviating the Hubble parameter tension~\cite{Escudero:2019gzq,Brinckmann:2020bcn,Lyu:2020lps}.


Due to its importance, the investigation of SNI has been rigorously pursued across multiple domains including cosmological, astrophysical, and laboratory experiments. 
Among them, various astrophysical and cosmological observational results prefer flavor non-universal secret interactions~\cite{Das:2020xke,Brinckmann:2020bcn}.
Moreover, the laboratory experiments have provided  stronger constraints on the SNIs with $\nu_e$ and $\nu_\mu$ than those with $\nu_\tau$~\cite{Burgess:1992dt,Lessa:2007up,Bauer:2018onh,Deppisch:2020sqh,Berryman:2022hds}.

Motivated by these observational and experimental preferences, we explore the scenarios where the SNIs are flavor non-universal and the mediators do not interact with the charged leptons in this paper.
Notably, we focus on the exciting potential of a variety of future tau neutrino experiments in directly probing the $\nu_\tau$ SNIs with less constraints from the laboratory experiments so far, compared to the other flavor SNIs.

For concreteness, in this paper, we adopt a light vector SNI scenario where a vector with the mass below 1 GeV couples exclusively to the SM active neutrinos, 
described by the term $\sum_{\alpha ,\beta} g_{\alpha \beta}Z'_\mu\bar{\nu}_\alpha \gamma^\mu \nu_\beta$, which provides Non-Standard Interactions (NSIs) involving light mediators \cite{Machado:2016fwi, Farzan:2015hkd, Farzan:2016wym}.
A viable scenario addressing these NSIs is proposed in Ref.~\cite{Farzan:2016wym}, which is  briefly explained in the next section. 
The light
$Z^\prime$ can be produced 
via three-body rare decays of pseudoscalar
mesons, if kinematically available
, which can be important decay channels due to no chiral suppression.
Sensitivities of meson decay experiments to secret couplings of neutrinos to $Z^\prime$  is studied in Ref.~\cite{Bakhti:2017jhm}. 
We note that while meson decay experiments are sensitive to the {\it sum} of the coupling strength squares involving charged leptons produced in the decay of charged mesons, i.e., $\sum_{\alpha \in \{ e,\mu,\tau \}} |g_{e \alpha}|^2$ and $\sum_{\alpha \in \{ e,\mu,\tau  \}} |g_{\mu \alpha}|^2$ (by identifying the produced charged lepton), neutrino detectors can detect produced neutrinos and are sensitive to {\it each} of the couplings $g_{e \alpha}$, $g_{\mu \alpha}$ and $g_{\tau \alpha}$. 
In order to obtain conservative sensitivities, we further assume that both of the production and the decay of $Z'$ are controlled by a single SNI parameter $g_{\alpha \beta}$.
In particular, we discuss  the possibility of
using the tau neutrino flux measurement to constrain the coupling of neutrinos with the new light gauge boson.
The upcoming neutrino detectors can benefit from their capability to detect high-energy neutrinos produced from heavy mesons. Additionally, their abilities of detecting $\nu_\tau$ directly can provide superb sensitivities on the NSI couplings~\cite{Bakhti:2022axo}.

As laboratory experiments probing beam-produced neutrinos, we adopt Forward Liquid Argon Experiment (FLArE100), SND$@$LHC, FASER$\nu2$, and SHiP for reference. 
The FASER \cite{FASER:2019aik}, FASER$\nu$  \cite{FASER:2021mtu}, and SND$@$LHC \cite{SHiP:2020sos} detectors 
are currently under operations 
in the tunnels located in the beam forward direction nearby ATLAS and have recently announced their first phase data~\cite{FASER:2023zcr,SNDLHC:2023pun}, which has opened up an era of intensity frontier searches for BSM at the LHC. 
To succeed these experiments, the Forward Physics Facility (FPF) which aims to host the next generation experiments during the running of the High Luminosity LHC (HL-LHC) is proposed~\cite{Feng:2022inv}.

The FPF neutrino experiments are expected to detect high number of neutrino interactions at the highest energies ever achieved. Thus, their measurements are crucial to
uncover neutrino interactions at energies above $\mathcal O (100\,{\rm GeV})$. 
The proposed FLArE, a liquid argon time projection chamber (LArTPC) located  at FPF as well as  FASER$\nu$2 are designed to detect millions of neutrino interactions, including tau neutrinos, and to search for long-lived BSM particles or dark matter. 
SHiP is an intensity-frontier beam dump proposed experiment which  aims to explore the domain of  weakly interacting hidden light  particles with
masses in the MeV- GeV range.
Sensitivity of the currently running FASER$\nu$ to secret neutrino interaction is previously studied in Ref.~\cite{Bahraminasr:2020ssz}.

In addition to the laboratory experiments, we consider Deep Underground Neutrino Experiment (DUNE) far detector to analyze the atmospheric data in probing SNI for the first time. 
We place special emphasis on the sensitivity of DUNE atmospheric data to SNIs 
by probing the appearance of {\it downward-going} $\nu_\tau$ and discuss how this strategy is strong, as well as providing valuable insights on the flavor structure of $Z^\prime$.
Note that DUNE will have both far detector, probing high energy atmospheric neutrinos, and near detector for high intensity lower energy neutrinos; we can expect the interplay between those in probing new interactions of neutrinos. 
Moreover, the two detectors' excellent event reconstruction, angular resolution, and abilities in identifying $\nu_\tau$ directly with track reconstruction can provide key information on SNI.
The potential of DUNE near detector (ND) to constrain the new interaction is studied in  Ref.~\cite{Bakhti:2018avv}. 
Similar study on $\nu_\tau$ appearance by a neutrino-philic mediator including those for the SNI in short-baseline laboratory experiments is recently proceeded in Ref.~\cite{Dev:2023rqb}.

This paper is organized as follows. 
In Sec.~\ref{sec:model}, we explain our simplified set-up from which useful analytic formulas are obtained.
In Sec.~\ref{sec:analysis}, we summarize the experimental details of our reference tau neutrino experiments and discuss the analysis strategies.
In Sec.~\ref{sec:results}, we then show our analysis results for each coupling $g_{\alpha \beta}$ while turning off the others, Finally, we conclude our results and leave discussion in Sec.~\ref{sec:conclusions}.


\section{Theoretical set-up}
\label{sec:model}

In this section, we explain our simplified set-up of vector SNI with a new sub-GeV mass vector boson $Z^\prime$.
The relevant effective Lagrangian includes:
\begin{align}
\mathcal L \supset \sum_{\alpha ,\beta} g_{\alpha \beta}Z'_\mu\bar{\nu}_\alpha \gamma^\mu \nu_\beta\,,
\end{align}
where $g_{\alpha \beta}$ represents the coupling between the new light boson $Z^\prime$ and neutrinos of flavor $\alpha$ and $\beta$, 
which does not have to be flavor-diagonal
. This interaction can lead to a new decay mode of meson to lepton, neutrino, and $Z^\prime$, which is followed by a subsequent $Z^\prime$ decay.~\footnote{ Note that the off-shell $Z^\prime$ production leads to a four-body decay process for $Z'$ heavier than the mother meson. Due to the extra phase space suppression, we do not consider this contribution here.}

Note that this interaction can arise from gauging different combinations of baryon number and lepton flavor/number~\cite{He:1990pn,Allanach:2018vjg}. However, the coupling of the electron to the new gauge boson $Z^\prime$ is subject to stringent constraints across a wide range of $Z^\prime$ mass and hence our models of interest should suppress the sizable couplings to the SM charged leptons.
A possible scenario giving rise to this interaction can be obtained from adopting a new gauge symmetry U(1)$^\prime$ along with a SM singlet heavy fermion $\Psi$ and a scalar particle both of which are charged under U(1)$^\prime$~\cite{Farzan:2016wym,Farzan:2017xzy}. 
Then the active neutrinos can couple to $Z^\prime$ by mixing with $\Psi$ when the new scalar particle is either a SM singlet or doublet; the active neutrinos of flavor $\nu_\alpha$ can be written as a linear combination of mass eigenstates $\nu_i$, ($i=1,2,3,4$): 
\begin{align}
\nu_\alpha =\sum_{i=1}^4 U_{\alpha i} \nu_i\,,
\end{align}
where $\nu_4$ is the heaviest mass eigenstate that gives the main contribution to $\Psi$. Integrating out the heavy fourth state, the light active neutrinos receive a coupling of the form $g_{\beta\alpha } Z_\mu' \bar{\nu}_\beta \gamma^\mu \nu_\alpha $
where $
 g_{\beta\alpha }=  g_\Psi U_{\alpha 4}U_{\beta 4}^*$ and $g_\Psi$ being the U(1)$^\prime$ gauge couping of $\Psi$. 
Note that a kinetic mixing between U(1)$^\prime$ and U(1)$_{\rm Y}$ can generically arise. 
Hence additional theoretical set-up should be assumed in such a way that the tree level mixing is turned off and the loop level mixing is induced by very heavy particles.

\begin{figure}[h]
\includegraphics[scale=0.26]{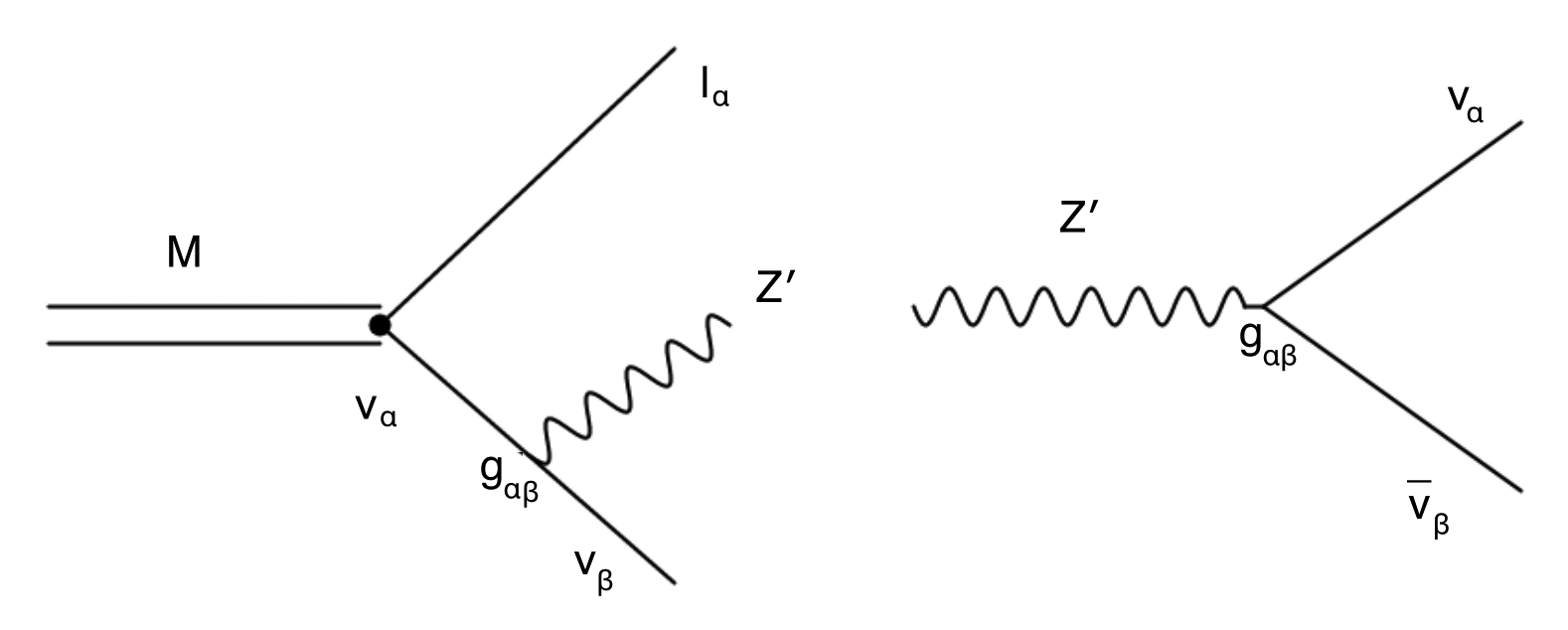}
\caption{Three body meson decay $M \rightarrow l \nu Z^\prime $ and subsequent $Z^\prime $ decays into a pair of a neutrino and an anti-neutrino.
\label{fig:process}
}
\end{figure}

Our process of interest is depicted in Fig.~\ref{fig:process}. 
In order to focus on the SNI from our reference scenario, we do not consider the baryonic couplings of $Z'$ throughout the whole analysis.
The flux of neutrinos in the lab frame coming from meson decay, 
$M \rightarrow Z' \nu \ell$ with subsequent decay of $Z' \rightarrow \nu \bar{\nu}$, is given by the equation:
\begin{align}\label{Eq.nuflux}
\Phi(E_\nu)= \frac{1}{4 \pi L^2} \int_{E_M^{min}}^{E_M^{max}}dE_{M} P_M(E_M) \left(\frac{dN_\nu}{dE_\nu}\right)_{lab} \frac{d \Omega_{r.M}}{d \Omega_{lab}}\,.
\end{align}
Here, $L$ represents the distance from the source to the detector and $P_M(E_M)$ is the rate of the meson injection in the lab frame.

The spectrum of the neutrino in the lab frame from the decay of a meson with an energy of $E_M$ is expressed as $\left(\frac{dN_\nu}{dE_\nu}\right)_{lab}$, which is related to the spectrum of neutrinos in the rest frame of meson, $dN_\nu/dE_\nu|_{r.M}$, as
\begin{align}
\left(\frac{dN_\nu}{dE_\nu}\right)_{lab}=\left.\left(\frac{dN_\nu}{dE_{\nu}}\right)\right|_{r.M} \frac{ \partial E_\nu|_{r.M}}{ \partial E_\nu|_{lab}}\,.
\end{align}

The value of $\left. E_\nu \right|_{lab}$ can be simply obtained by kinematics. If the direction of a neutrino reaching the detector coincides with that of the spatial momentum of the meson beam, for instance, we can write $E_\nu|_{lab}=E_\nu|_{r.M}(1+v_M)\gamma_M$, in which $v_M$ is the meson velocity in the lab frame and $\gamma_M=(1-v_M^2)^{-1/2}$, therefore  $\frac{\partial E_\nu|_{r.M}}{ \partial E_\nu |_{lab}}=\gamma_M (1-v_M).$
Note that $d \Omega_{r.M}/d\Omega_{lab}$ takes care of focusing of the beam in the direction of the detector and is given by $(1+v_M)/(4(1-v_M)) \simeq \gamma_M^2$
and $\left.\left(\frac{dN_\nu}{dE_{\nu}}\right)\right|_{r.M} $ is the total spectrum of the  (anti-)neutrino produced from both meson and $Z'$ decay:
\begin{align}\label{Eq.anuspec}
\left.\left(\frac{dN_\nu}{dE_\nu}\right)\right|_{r.M}= \left.\left(\frac{dN_\nu}{dE_\nu }\right)\right|_{r.M}^{Z' decay}
+ \frac{N_0}{ \Gamma (M\longrightarrow  l \nu Z')}\frac{d\Gamma ( M\longrightarrow l \nu Z')}{dE_{\nu}} \,,
\end{align}

where $N_0$ is the total number of the neutrinos produced from the $M^+$ ($M^-$)
decay, which is $\nu_\alpha$ ($\bar \nu_\alpha$) in the left panel of Fig.~\ref{fig:process}. 
For the electron decay mode $M \to e \nu Z'$, we neglect the mass of electron and obtain the decay rate analytically, while the values for the heavier leptons, $M \to \mu \nu Z'$ and $M \to \tau \nu Z'$ are obtained numerically.

As previously stated, the presence of the new light gauge boson $Z'$ leads to an enhanced three-body decay rate of the 
pseudo scalar meson without chiral suppression compared to the two-body decay cases due to the longitudinal component of the new massive gauge boson. This enhancement arises from the new interaction between the new massive gauge boson and neutrinos, and the decay rate scales as $g_{\alpha \beta} ^2/m^2 _{Z^{\prime}}$, which results from the summation over the new gauge boson polarizations, i.e., $\Sigma_i \epsilon^{\mu} _i (k) {\epsilon^{\nu}}^{\star} _i (k)= -g^{\mu \nu} + k^\mu k^\nu / m^2 _{Z^{\prime}}$. This phenomenon is analogous to the $W$ boson emission in top quark decay ($t \rightarrow b W$), where the decay rate is proportional to $1/m^2 _{W}$. The enhancement in both cases arises from the polarization sum resulting from the spontaneously broken gauge symmetry.
For the decay modes into $e^\pm$, 
the differential decay rate with polarization perpendicular to the $Z'$ momentum $(\epsilon_1,\epsilon_2)$ and parallel to the $Z'$ momentum $(\epsilon_3)$ can be expressed as:
\begin{align}
\left.\frac{d\Gamma(M \longrightarrow e\nu_\alpha Z')}{dE_{Z'}} \right|_{1,2} &= \frac{ f_{M }^2 g_{e\alpha}^2 G_F^2 \cos^2 \left(\theta _C\right)}{96 \pi ^3 m_M}p_{Z'} \left(-2 E_{Z'} m_M+m_M^2+m_{Z'}^2\right)\,, \\
\left.\frac{d\Gamma(M\longrightarrow e\nu_\alpha Z')}{dE_{Z'}} \right|_{3} &= \frac{f_{M }^2 g_{e\alpha}^2 G_F^2 \cos^2 \left(\theta _C\right)}{96 \pi ^3 m_M  m_{Z'}^2}p_{Z'} \left(E_{Z'} m_M-m_{Z'}^2\right)^2\,,
\end{align}
where $f_M$ is the meson decay constant. 
Observing the decay into the longitudinal mode, we can see that it is proportional to $g_{e\alpha}^2/m_{Z'}^2$ and will be enhanced for $m_{Z'}\ll m_M$. For the case of pion which is dominantly produced in various accelerators, the total decay rate is given by:
\begin{align}
\begin{split}
\Gamma(\pi\longrightarrow e\nu_\alpha Z') &= \frac{g_{e\alpha}^2 G_F^2 \cos^2 \left(\theta _C\right) f_\pi^2}{6144 \pi ^3 m_\pi^3 m_{Z'}^2} \times \\
&\quad  \left(m_\pi^8+72 m_\pi^4 m_{Z'}^4-64 m_\pi^2 m_{Z'}^6+24 \left(3 m_\pi^4 m_{Z'}^4+4 m_\pi^2 m_{Z'}^6\right) \log \left(\frac{m_{Z'}}{m_\pi}\right)-9 m_{Z'}^8\right)\,.
\end{split}
\end{align}
\begin{figure}[h]
\includegraphics[scale=0.49]{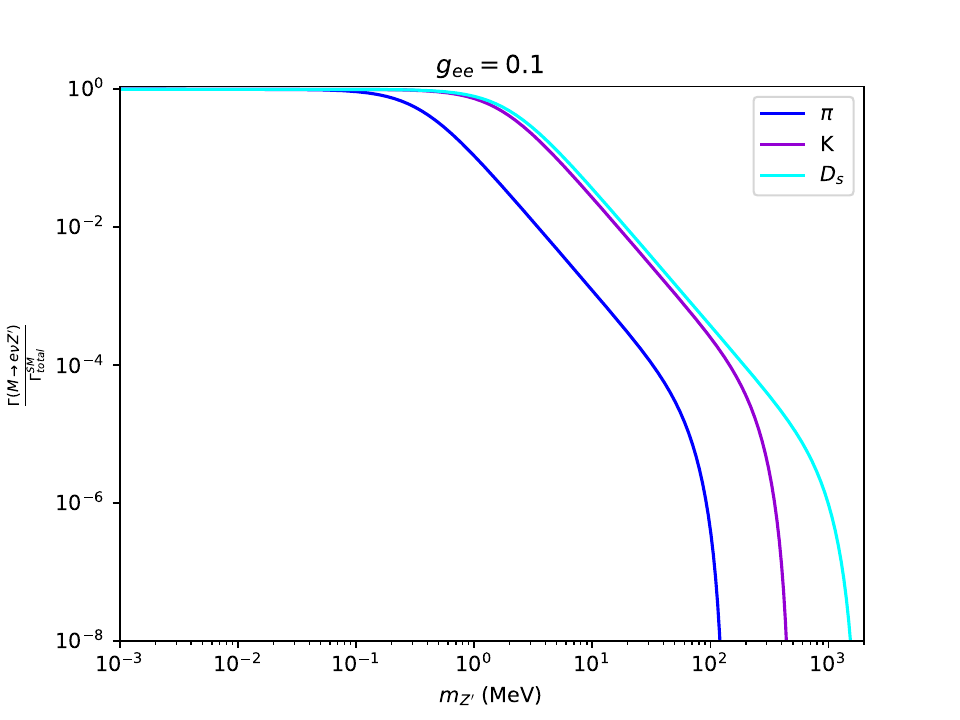}
\includegraphics[scale=0.49]{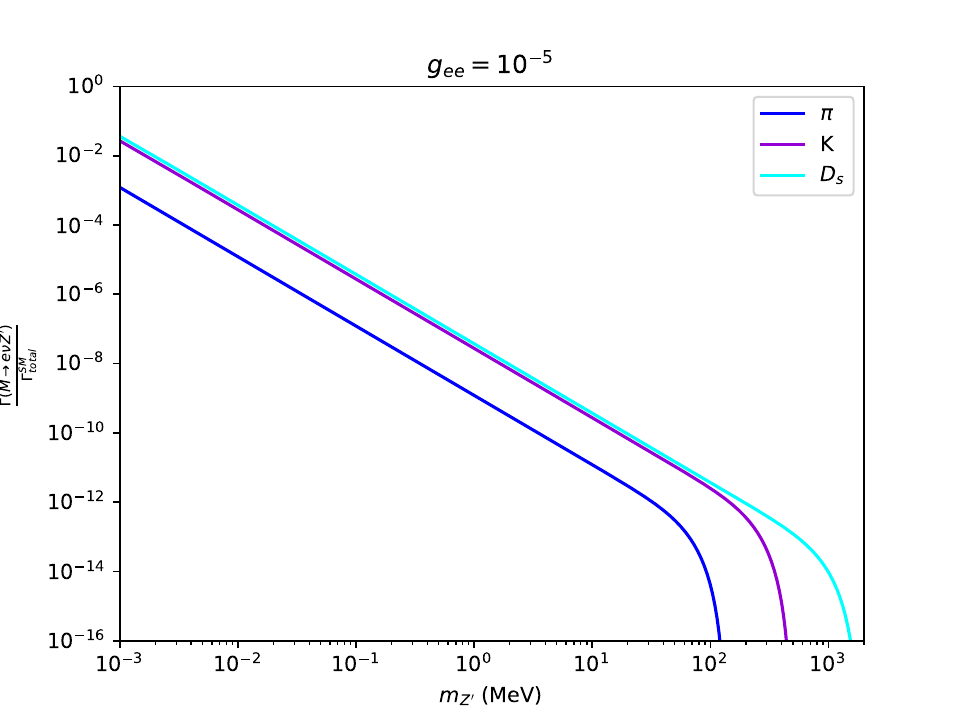}
\caption{The branching ratio of the meson three-body decay to electron, anti-neutrino and new gauge boson  $\frac{\Gamma(M \rightarrow e \nu Z^{\prime})}{\Gamma_{\text{total}}^{\text{SM}}}$ as a function of $m_{Z^\prime}$ for different mesons, namely $\pi$, $K$, and $D_s$. We have fixed the value of 
  $g_{ee}=0.1$ $(g_{ee}={10^{-5}})$ in the left (right) panel.}
\label{fig:ratio}
\end{figure}
Figure \ref{fig:ratio} shows the branching ratio of the meson's three-body decay to electron, anti-neutrino, and new gauge boson, $\left(\frac{\Gamma(M \rightarrow e \nu Z^{\prime})}{\Gamma_{\text{total}}^{\text{SM}}}\right)$,  as a function of $m_{Z^\prime}$ for various mesons, including $\pi$, $K$, and $D_s$ assuming $g_{ee}=0.1$ (left) and $g_{ee}={10^{-5}}$ (right). 
For large new gauge couplings of $\lesssim \mathcal O (0.1)$, we can clearly see that the three-body branching ratios can be dominant over the conventional chiral suppressed two-body decays and would be easily constrained by various experiments.
As one expects, the branching ratio decreases rapidly as the gauge boson mass approaches the mass of the charged meson which makes the process kinematically forbidden.

The $Z'$ gauge boson with masses of $\mathcal O ({\rm MeV} - 100\,{\rm MeV})$ subsequently decays into $\nu \bar{\nu}$ before reaching the detector
producing signals at neutrino detectors over a wide range of $g_{\alpha \beta}$.
The total decay rate of $Z'\longrightarrow\nu_\alpha\bar{\nu}_\beta$ for all the polarizations is given by:
\begin{align}
\Gamma(Z'\longrightarrow\nu_\alpha\bar{\nu}_\beta)=\frac{g^2_{\alpha\beta} m_{Z'}}{24\pi}\,.
\end{align}

The number of neutrinos originating from the decay of $Z^\prime$ particles before reaching the detector is given by
\begin{align}\label{Eq.NOFzpdecay}
N=N_0\left(1-e^{-\Gamma L / \gamma} \right)\,,
\end{align}
where $N_0$ is the number of produced $Z^\prime$, $L$ is the distance between the $Z^\prime$ production point and the detector, and $\gamma = E_{Z^\prime}/m_{Z^\prime}$ is the boost factor. In Eq.~(\ref{Eq.NOFzpdecay}), if $\Gamma L / \gamma \gg 1$, almost all $Z^\prime$ particles decay before reaching the detector.

Before closing this section, let us briefly comment on the cosmological effect of a light $Z'$. 
For $Z'$ with masses $m_{Z'} \lesssim \mathcal O ({\rm MeV})$, it is inevitable to consider its contribution to the radiation energy density without Boltzmann suppression at the time of Big Bang Nucleosynthesis (BBN).
In the presence of SNI, the new gauge boson can be generated via inverse decay $\nu+\bar{\nu}\rightarrow Z^\prime$ and neutrino-antineutrino annihilation $\nu+\bar{\nu}\rightarrow Z^\prime+Z^\prime$~\cite{Huang:2017egl}. 
The new gauge boson $Z'$ can contribute to the extra radiation species $\Delta N_{\rm eff}$ when it is in thermal equilibrium with active neutrinos around $T \sim 1~{\rm MeV}$ after neutrino decoupling era.~\footnote{This is for $Z'$ heavier than the active neutrinos and much lighter neutrino cases are also discussed in Refs.~\cite{Forastieri:2015paa,Berryman:2022hds}.}
As a conservative limit, we adopt the constraints on $\Delta N_{\rm eff}$ with 90 \% C.L., which is $\Delta N_{\rm eff} \lesssim 1$, from Ref.~\cite{Huang:2017egl} assuming a flavor universal SNI. 
The combined constraints with the abundances of the primordial elements are similar in the reference.
We expect the flavor non-universal and off-diagonal cases involving $\nu_\tau$ which is more proper to be applied in our analysis would provide weaker bounds but we do not pursue this direction here and leave more detailed study to a future work.
Hence we simply apply the nominal bound on the vector boson for $m_{Z'} \lesssim 5$ MeV from Ref.~\cite{Huang:2017egl}.  
Note that possible baryonic interactions of $Z'$ can provide extra constraints on the abundances of the primordial elements but we do not include those to focus on SNI here, as stated earlier.
Other scenarios such as scalar SNI can have weaker bounds due to the smaller degrees of freedom.

\section{Analysis strategies} 
\label{sec:analysis}

\subsection{Accelerator based Neutrino Experiments}

In this section we explain the details of the reference experiments and the analysis strategies.
The FLArE100, FASER$\nu$2, SND$@$LHC, and SHiP experiments allow us to probe the relevant parameter space for the relatively heavy $Z^\prime$ up to $\lesssim 1$ GeV since their beam energies are high enough to produce heavy mesons such as charmed mesons.
Moreover, these detectors have the potential to collect a large number of 
tau neutrino events and identify those, providing an opportunity to use tau neutrino flux measurements in probing possible new interactions of neutrinos.
To compute the number of events, we have taken the neutrino cross section and the energy spectra of the charged mesons from Ref.~\cite{FASER:2019dxq}.
We further assume the perfect energy resolutions and $80\%$  efficiencies for $\nu_e$ and  $\nu_\mu$ in the aforementioned experiments for simplicity.

The FPF is expected to host far-forward experiments such as FASER$\nu$2, a 20-ton emulsion
detector; Advanced SND@LHC (AdvSND), a successor to SND@LHC; and FLArE, a proposed liquid argon time projection chamber with an active volume of 100 tons~\cite{Feng:2022inv}. 
These experiments have potential to detect millions of TeV-energy neutrinos. 
The AdvSND features a 5-ton fiducial mass that represents a substantial increase of 6.25 times compared to the SND@LHC experiment.
Furthermore, it is expected to have the final integrated  luminosity 20 times higher than SND@LHC (150 fb$^{-1}$), resulting in a total 125 times larger data.
We have taken the details of the aforementioned experiments from Ref.~\cite{Kling:2021gos}.

The Search for Hidden Particles  (SHiP) experiment is a proposal of
fixed target facility at the CERN Super Proton Synchrotron (SPS) which aims to search for 
light BSM particles with tiny interactions with the SM particles avoiding the experimental constraints thus far, so called hidden particles~\cite{SHiP:2021nfo}.
The other main purpose of SHiP is directly observing  $\nu_\tau$ 
and $\bar \nu_\tau$
events. 
Benefiting from high statistics it can perform active neutrino physics. 
Inside SHiP, a detector called SND$@$SHiP will be installed for the study of active neutrino cross-sections and angular distributions. This is expected to be located about 46 m behind the interaction point and 
detect about 12000 $\nu_{\tau}$ events within 5 years of operation, which is quite large compared to FASER$\nu$ experiment detecting about 11 $\nu_{\tau}$ events by 2023.
Interestingly, SHiP hosts a hadron absorber that  light mesons such as pions or Kaons can interact with before decaying to neutrinos.
Hence, the fraction of the charmed meson increases compared to the lighter mesons.
Since the $D_s$ meson decays are the main sources of tau neutrino production at the SPS with the beam energy 400 GeV, it is possible to have a large tau neutrino flux. 
Moreover, SHiP will have the opportunity to observe  the tau anti-neutrino for the first time and perform its cross section measurements.
In this respect, SND@SHiP will also be an excellent experiment searching for the BSM particles interacting with tau neutrinos.

\subsection{Atmospheric Neutrino Experiments}

In addition to the beam produced neutrinos, atmospheric neutrinos can be used to set stringent bounds on the new couplings. Although the number of charmed mesons produced in the atmosphere is smaller than
those in the aforementioned accelerator experiments, it is possible to effectively probe tau neutrinos by reducing most of the backgrounds remarkably.
We adopt DUNE far detector to confirm our expectation.

All of the current neutrino  oscillation experiments explore $\nu_\mu$ and $\nu_e$ disappearance and $\nu_\mu \rightarrow \nu_e$
appearance channels (plus anti neutrino channels). 
Both atmospheric and neutrino beam experiments have confirmed 
the $\nu_\mu \rightarrow \nu_\tau$ oscillation by the disappearance of $\nu_\mu$.
This is because the reconstruction and identification of $\nu_\tau$ events pose significant challenges due to the prompt and semi-visible decay of the $\tau$ leptons. 
In particular, the misidentified neutral current (NC) scattering of any flavor neutrinos can mimic the $\tau$ lepton signal and is hard to be rejected.
Moreover, the energy threshold to detect the 
charged current (CC) scattering of $\nu_\tau$ off the matter producing $\tau$ lepton is as high as
$E_{\nu_{\tau}} \gtrsim 3.35$ GeV for the nuclear scattering process and $E_{\nu_{\tau}} \gtrsim 3.1$ GeV  for the electron scattering process, which are mostly beyond the reach of the current beam neutrino experiments. 

On the other hand, DUNE is expected to have capabilities of identifying and reconstructing the $\tau$ lepton signals due to the characteristic of the Liquid Argon Time Projection Chamber (LArTPC) detector with an excellent position resolution~\cite{DeGouvea:2019kea,Machado:2020yxl}. 
In particular, the Long-Baseline Neutrino Facility (LBNF) will be equipped with the 120 GeV Neutrinos at the Main Injector (NuMI) beam providing the center-of-mass energies well above 3 GeV to observe the $\nu_\tau$ CC processes and the near detector complex will host a variety of detectors which can reduce the backgrounds~\cite{DUNE:2021tad}.

In LBNF, the first oscillation maxima for DUNE occurs around 2.5 GeV which is below but very close to the tau neutrino detection energy threshold. This will cause some ambiguity in the measurement of $\Delta m^2 _{31}$ and $\sin^2 \theta_{23}$~\cite{DeGouvea:2019kea}. 
By comparison, DUNE far detector covering a wide range of $L/E$ and benefiting from large flux can provide a promising tool to search for $\nu_{\tau}$.  Hence 
atmospheric data can provide a clearer
first oscillation maxima together with 
an excellent angular resolution (zenith angle resolution is $\sim 5^\circ$ for $\nu_\tau$ CC and $\sim 7^\circ$ for NC) and energy resolution ~\cite{MammenAbraham:2022xoc}, making the atmospheric data advantageous in more accurate measurements of oscillation parameters.

Notice that upward-going atmospheric neutrinos travelling through a larger baseline can effectively oscillate into 
$\nu_\tau$. On the other hand, we do not expect $\nu_\tau$ signals within the standard oscillation model for downward-going atmospheric neutrinos since their baselines are too short to oscillate.
Therefore, the unexpected {\it downward-going $\nu_\tau$ appearance} will be a unique signal of non-standard interaction without suffering from large background contamination in the atmospheric neutrino experiments such as DUNE far detector. 
In addition, the far detector of DUNE has much larger fiducial volume than those in the accelerator based experiments, increasing its capabilities in searching for non-standard interactions.
Note that similar expectation of the unexpected $\nu_\tau$ appearance in the beam produced short-baseline neutrino experiments is studied in Ref.~\cite{Dev:2023rqb}. In this study we will explore how atmospheric neutrino data of DUNE can provide wonderful sensitivities in probing the SNI.
The details of the experiments we have used are given in table \ref{table}.

\begin{table}
\begin{tabular} { | p {3 cm} | p {3 cm} | p {3 cm} | p {3 cm} |p {3 cm} |} 

    \hline
    \multicolumn{5} { | c | }{Detector ~~~~~~~~~~~~~~~~~~~~~~~~~~~~~~~~~~number of events}\\
    \hline
    Detector name & mass & $\nu_{e} + \bar{\nu}_{e}$ & $\nu_{\mu} + \bar{\nu}_{\mu}$ & $\nu_{\tau} + \bar{\nu}_{\tau}$ \\
    \hline
    SND@LHC  & 800 kg & 250  & 1000  & 11  \\
   FASER$\nu$2   & 20 tonnes & $7.5 \times 10^4$   & 4 $\times 10^5$  & $1.7 \times 10^3$  \\
    FLArE100  & 100 tonnes & $2.5 \times 10^4$  & $1.38 \times 10^5$  & $1.3 \times 10^4$  \\
   SHiP & 10 tonnes & $3.4 \times 10^4$  &  $2.35 \times 10^5$ & $1.2 \times 10^4$  \\
  DUNE & 40 kilo-tonnes &  $1.6 \times 10^4$ &  $2.4 \times 10^4$  & $150$  \\
    \hline 
    \end{tabular}  
    \caption{\label{table} Estimated numbers of standard model neutrino events assuming a final integrated luminosity of 150 fb$^{-1}$ for SND@LHC, while 3000 fb$^{-1}$ for FASER$\nu2$ and FLArE100. For SHiP, we assume $2 \times 10^{20}$ POT in five years. We assume a data-taking period of 10 years for DUNE atmospheric neutrinos.
    }
    
\end{table}

 ~~~~~~
 ~~~~~~~~~~~~~~~~~~

 ~~~~~~~~~
 ~~~~~~~~~~~

\section{Results}
\label{sec:results}

\begin{figure}
 \includegraphics[scale=0.75]{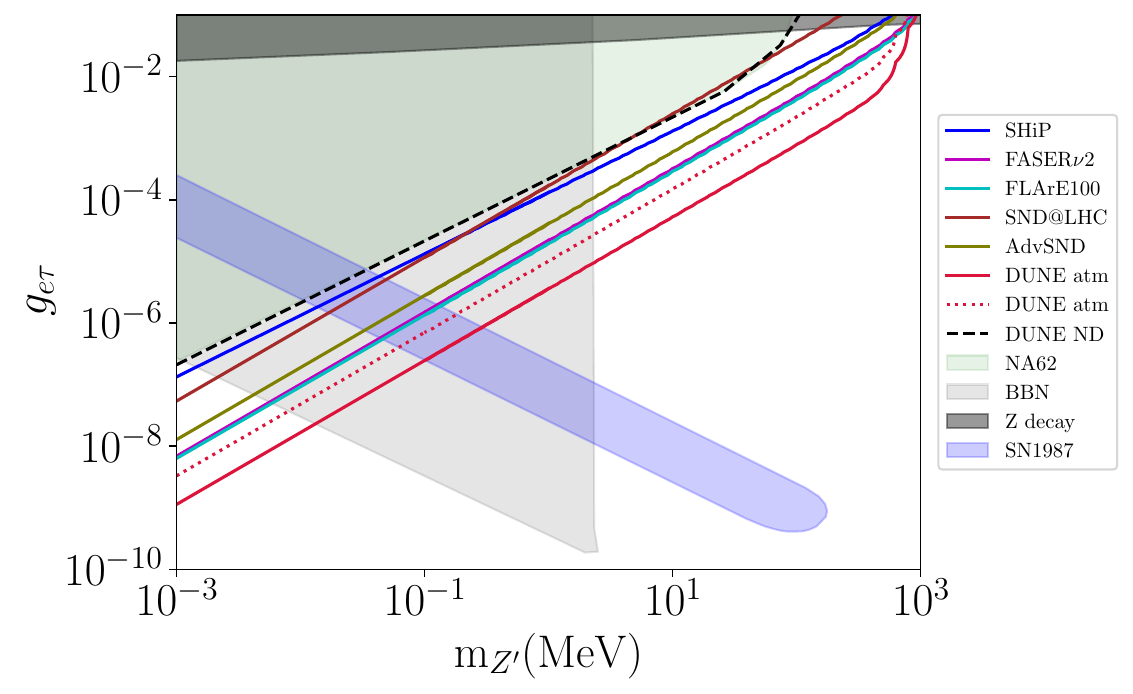}
 \caption{ The upper bound on $g_{e\tau}$ vs. $m_{Z^\prime}$ at 90\% C.L.. The red curve corresponds to the sensitivity of DUNE with the ten years of atmospheric data assuming no background. The dotted red curve corresponds to the DUNE atmospheric data taking into account the background by NC scatterings of any flavor neutrinos estimated in Ref.~\cite{aurisano2021experimental}, which is 70 events for 10 years. 
The cyan and purple curves show the sensitivity of FLAE100  and FASER$\nu2$ to constrain $g_{e\tau}$, respectively. The blue and  curve corresponds to the  SHiP, i.e., SND@SHiP experiment and the green one corresponds to the  Advanced SND$@$LHC experiments.  The dashed black curve  shows the bound from DUNE near detector \cite{Bakhti:2018avv}.  The gray region shows the BBN constraint \cite{Huang:2017egl}. The dark gray and light green regions shows the current constraint from $Z$ boson decay and NA62, respectively \cite{Laha:2013xua,NA62:2012lny}. 
The light blue region indicates the constraint from core collapse supernova  \cite{Akita:2022etk}.}\label{get}
 \end{figure}

In this section, we show our analysis results by displaying the expected sensitivities of various reference experiments in the plane of $g_{\alpha \beta} - m_{Z'}$ along with the current constraints.
In all of our results, we assume a chosen $g_{\alpha \beta}$ is the only non-zero SNI coupling to make the analysis conservative and simple. 
From observing the $\nu_\tau$ events, the SNI couplings $g_{\alpha \tau}$ can be directly probed. 

The sensitivities on $g_{\tau \tau}$ can be, in principle, dominantly obtained from the process of $D_s \to \tau \nu_\tau Z' \to \tau 3 \nu_\tau$. 
However, we expect those are very weak due to the small flux of $D_s$ and the phase space suppression for a given mass of $Z'$. 
On the other hand, $g_{e\tau}$ and $g_{\mu\tau}$ can be probed in the processes $D_s \to e \nu_e Z'$ and $D_s \to \mu \nu_\mu Z'$, respectively, again without chiral suppression compared to $D_s \to \tau \nu_\tau Z'$, providing more phase space. In addition, those SNIs can be probed from the lighter meson decays.
We estimated that BR($D_s \to \tau \nu_\tau Z'$) for $g_{\tau \tau}=0.1$ and $m_{Z^\prime} \sim 10$ MeV is about $10^{-4}$ times smaller than BR($D_s \to e \nu_e Z'$) in Fig.~\ref{fig:ratio}.

In order to show the effectiveness of our analysis strategies observing the $\nu_\tau$ events, we analyze the  other SNI couplings, i.e., with $\nu_e$ or $\nu_\mu$ but not $\nu_\tau$.
Note that our $Z'$ from the reference models might be further constrained by its baryonic interactions but we do not include such a possibility as a conservative approach.

Figure \ref{get} displays the $90\%$ C.L. current constraints and future sensitivities
on $g_{e\tau}$ versus $m_{Z^{\prime}}$, while all the other SNI couplings are set to zero.
The analysis takes into account all meson decays, including $\pi$, $K$, and $D_s$ to leptons and neutrinos ($\pi , K , D_s \rightarrow l , \nu $). The green region indicates the current exclusion limit from NA62 \cite{NA62:2012lny}, while the dark and light gray regions represent the constraints from $Z$ boson decay and BBN, respectively \cite{Laha:2013xua,Huang:2017egl}. 
Note that the late decay of $Z' \to \nu_\alpha \nu_\beta$ prior to the recombination epoch can possibly contribute to extra $\Delta N_{\rm eff}$ from the observation of Cosmic Microwave Background (CMB), which will be stronger than the currently displayed BBN bound.
However, the detailed fitting of the CMB data in the presence of flavor non-universal and off-diagonal SNI is nontrivial and hence we leave more dedicated study to a future work without displaying those bounds throughout this work.
We can also apply a bound from  
the observation of the power spectrum of CMB due to the late neutrino free streaming~\cite{Archidiacono:2013dua, Das:2020xke, Berryman:2022hds} but it is far weaker than that from NA62 for a simple universal couplings case, $g_{ee} = g_{\mu \mu} = g_{\tau \tau}$.
We hence do not show the CMB power spectrum constraint here.
The light blue region indicates the constraint from core-collapse supernovae \cite{Akita:2022etk}. 

The plot  demonstrates that FLArE100 (cyan curve) and FASER$\nu$2 (purple curve) can set comparable and the most stringent constraints on $g_{e\tau}$ among future beam experiments. 
Note that FLArE100 has the largest fiducial volume with the background events comparable to the much smaller detectors, SND@SHiP or AdvSND, as can be seen in Table~\ref{table}. 
Also, FASER$\nu2$ has a smaller fiducial volume but with much smaller number of backgrounds even compared to FLArE100. 
With the difference in the shape of the neutrino flux, the above advantages make FLArE100 and FASER$\nu$2 promising in probing the $g_{e\tau}$ coupling.
For $m_{Z^\prime} \gtrsim {\rm few}$ MeV, they can improve the current constraint by about one order of magnitude, and by more than one order of magnitude for $m_{Z^\prime} \lesssim$ 60 keV. Additionally, the plot shows that SHiP is more sensitive to the new coupling above a few MeV compared to SND$@$LHC. This is due to the fact that the fraction of the produced charm mesons to lighter mesons at SHiP experiment is higher than SND$@$LHC
due to the presence of hadron absorber.
It is important to note that our analysis focuses solely on the far detector, where we expect a similar neutrino flux as SND@LHC. However, it is worth mentioning that incorporating the near detector may potentially lead to significant improvements in the obtained results. As depicted in the plot, Advanced SND@LHC demonstrates a comparable sensitivity on the parameter $g_{e \tau}$ when compared to the FASER$\nu2$ and FLArE100 experiments.

It is remarkable that the atmospheric data at DUNE with 10 years of running can provide most stringent sensitivities for $m_{Z'} \gtrsim 1$ MeV and $m_{Z'} \lesssim 60$ keV. Note that the direction of $\nu_\tau$ is crucial, i.e., the downward-going events determine the sensitivities. 
The red solid line corresponds to the zero-background assumption while the red dotted line shows the inclusions of the background by the NC scatterings of any flavor downward-going neutrinos expected in Ref.~\cite{aurisano2021experimental}, which is 70 events for 10 years.
In performing our analysis, we use the Honda atmospheric neutrino flux model~\cite{Honda:2004yz}. 
We assume $100 \%$ efficiency for $\nu_\tau$ event reconstruction
for simplicity, following the relevant study~\cite{Machado:2020yxl}.
Notice that the efficiency for reconstructing tau neutrino tracks in DUNE atmospheric data depends on various factors such as the energy, direction of the neutrino, the properties of the detector, and the reconstruction algorithms used. Nevertheless, DUNE is designed to have excellent spatial and angular resolutions. 
The detector consists of a large volume of liquid argon, which allows for precise tracking and energy measurements of particles produced in neutrino interactions. The detector is also complemented by a highly sophisticated software system for event reconstruction, which is continually being improved to increase the efficiency and accuracy of tau neutrino reconstruction. We assume angular resolution ($\Theta_{zen}$ resolution) of $5^\circ$ for $\nu_\tau$ CC \cite{aurisano2021experimental}.
More dedicated study in this direction in collaboration with the experimentalists is also possible in the future.
Although limited, the DUNE near detector (ND) can also play roles in observing the SNI, which is shown as black dashed line following Ref.~\cite{Bakhti:2018avv}.
Note that these experiments can provide the searches for SNIs complementary to cosmological (gray) and astrophysical (blue) probes. 
In some parameter regions such as $m_{Z'} \gtrsim \mathcal O ({\rm MeV})$ or $g_{e \tau} \lesssim \mathcal O (10^{-7})$ while $m_{Z'} \lesssim \mathcal O (10^{-2}\,{\rm MeV})$, the ground based experiments exhibit better sensitivities.

\begin{figure}[h]
 \includegraphics[scale=0.75]{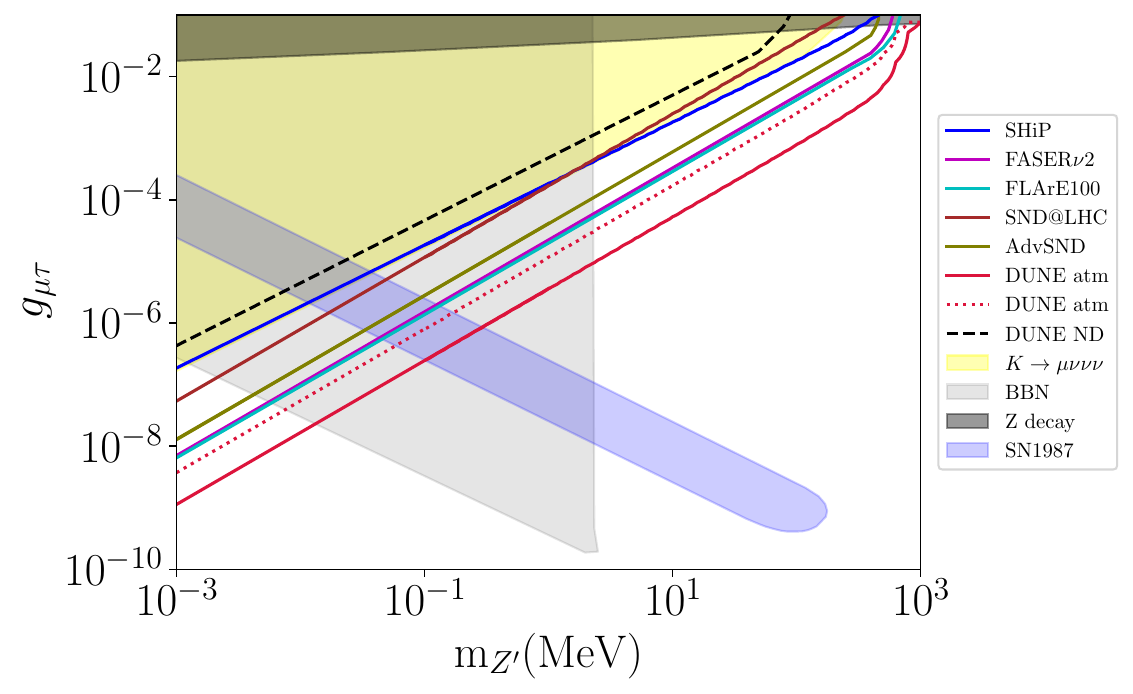}
 \caption{ The upper bound on $g_{\mu \tau}$ vs. $m_{Z^\prime}$ at 90\% C.L..
 The yellow region corresponds to the current constraint from $K \rightarrow \mu \nu \nu \nu$ \cite{E949:2016gnh}  } 
  \label{gmt}
\end{figure}
 
\begin{figure}
 \includegraphics[scale=0.75]{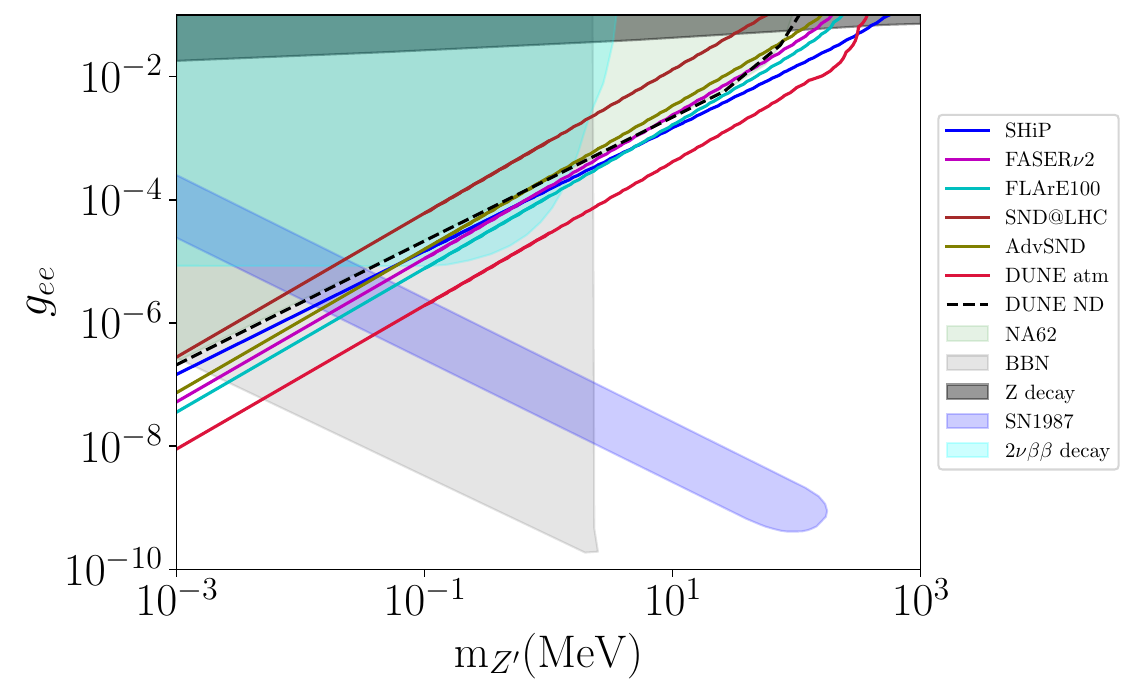}
 \caption{The upper bound on $g_{ee}$ vs. $m_{Z^\prime}$ at 90\% C.L..
 }\label{gee}
\end{figure}

\begin{figure}
 \includegraphics[scale=0.75]{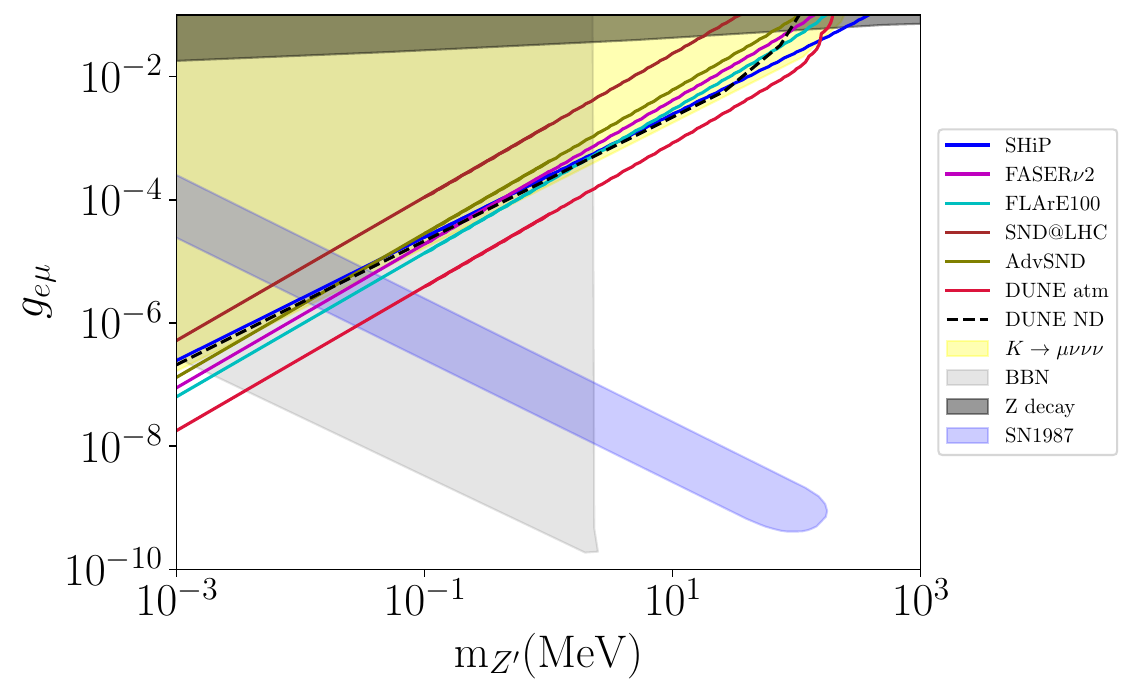}
 \caption{The upper bound on $g_{e \mu}$ vs. $m_{Z^\prime}$ at 90\% C.L..}\label{gem}
 \end{figure}

Figure \ref{gmt} displays the upper bound at $90\%$ C.L. on $g_{\mu \tau}$ as a function of $m_{Z^{\prime}}$. As observed in the plot, FLArE100 (cyan curve) and FASER$\nu$2 (purple curve) can improve the current constraint on $g_{\mu \tau}$ by more than one order of magnitude for $M_{Z^\prime}> {\rm few}$ MeV and for $M_{Z^\prime} <$ 60 keV. Furthermore, SHiP can slightly enhance the current constraint on $g_{\mu \tau}$ for masses larger than a few MeV, whereas SND$@$LHC can slightly improve the current bound for masses lower than few keV.
Note that the NA62 bound is not applied here since $g_{\mu \tau}$ is the coupling between $\nu_\mu$ and $\nu_\tau$.
Instead, we applied the experimental constraints from $K \to \mu \nu_\mu$ since the new gauge boson can be produced from $\nu_\mu$ and produce $K \to \mu \nu_\tau \nu_\mu \nu_\tau $ \cite{E949:2016gnh}. The corresponding bound is expressed as the yellow shaded region.
Again, we expect the DUNE atmospheric data with the 10 years of running can provide the best sensitivity.

For comparison, we include the analysis for $g_{ee}$ and $g_{e\mu}$ in Fig.~\ref{gee} and \ref{gem}, respectively.
We can clearly see the sensitivities are weaker by about an order of magnitude than those in Figs.~\ref{get} and \ref{gmt} due to the background contamination for the other flavors of neutrino events.
We have taken the SM events as the background for $\nu_e$ and $\nu_\mu$. The number of $\nu_e + \bar{\nu}_e$ is $1.6 \times 10^4$. Also the number of $\nu_\mu + \bar{\nu}_\mu$ is $2.4 \times 10^4$.  Note that the shape of the spectrum is important for $g_{ee}$ and $g_{e\mu}$  and the direction of the neutrinos is non-critical.

For $m_{Z^\prime} <$ few keV, FLArE100 can improve the current constraint on $g_{ee}$, while SHiP shows better sensitivities for $m_{Z'} \gtrsim 400$ MeV due to its 
higher sensitivity to neutrinos originating from heavy meson decays, such as $D_s$. These results highlight the importance of studying heavy meson decays to further constrain the coupling of secret neutrino interactions in the higher $Z'$ mass region.

Interestingly, the atmospheric neutrino data in the 10 years of running of the DUNE far detector can still provide excellent sensitivities better than the accelerator experiments in most of the parameter space of $m_{Z'} \lesssim 500$ MeV due to the size of the fiducial volume
and the shape of the flux.
As the mass of $Z'$ approaches to GeV level, SHiP can be more sensitive since the flux of heavy mesons in the atmosphere decreases while the large backgrounds of $\nu_\mu$ or $\nu_e$ are still not effectively rejected.

In addition, the SNI with $g_{ee}$ can induce the two-neutrino double beta decay ($2\nu \beta \beta$) which is also expected in the SM, even without lepton number violating interactions~\cite{Deppisch:2020sqh}. 
However, the current bound (cyan shaded region in Fig.~\ref{gee}) is still weaker than the combined constraints from BBN and NA62.
We expect the sensitivities on $g_{e \mu}$ are even weaker that those on $g_{ee}$ because the production rate of $\nu_\mu$ in both atmospheric and accelerator data is higher.

As can be observed from Fig.~\ref{get}$-$\ref{gem} atmospheric data of DUNE can set the most stringent bound on the new coupling among the reference experiments. As indicated in Fig.~\ref{get}, atmospheric data is the most sensitive probe on $g_{e\tau}$ even after including the NC background. Notice that, in our analysis for atmospheric neutrinos, we have fixed the flux value and did not include the uncertainty of the shape of the flux. The obtained sensitivities
on $g_{ee}$ and $g_{e \mu}$ can be modified significantly including this uncertainty. 
On the other hand, the expected sensitivities on  $g_{e \tau}$ and $g_{\mu \tau}$ are quiet robust with respect to the flux uncertainty since the standard interactions do not produce downward-going $\nu_\tau$.
We emphasize again that these two couplings are sensitive to the direction of the tau neutrino and the shape of the background. 

Finally, it is fair to leave a comment that our $Z'$ can induce lepton flavor violating rare decays such as $\mu \to e \gamma$ in the two loop level.
However, our naive estimation shows the contribution can be negligible for the couplings below $g_{\alpha \beta} \lesssim \mathcal O (10^{-2})$ compared to the experimental limits so far.
More exact calculation is beyond the scope of this paper and we leave the related study to a future work.


\section{CONCLUSION}
\label{sec:conclusions}

The upcoming beam and atmospheric tau neutrino experiments offer a promising avenue to explore the hidden interactions between neutrinos, whose identification is highly crucial in various fields, including neutrino physics, dark matter physics, grand unified theories, astrophysics, and cosmology.
For concreteness we adopted a scenario with a light gauge boson $Z^\prime$ with coupling $g_{\alpha \beta}$ to $\nu_\alpha$ and $\nu_\beta$. 
Our analysis highlights the importance of DUNE atmospheric data in  
obtaining the best sensitivities 
on $g_{\alpha \beta}$ for the $1~{\rm MeV} \lesssim m_{Z^\prime} \lesssim 500~{\rm MeV}$ mass range as well as for 
$m_{Z^\prime} \lesssim \mathcal O ({\rm keV})$, 
with the potential to improve the current constraint by up to two orders of magnitude. Notice that we have assumed angular resolution of $5^\circ$ to perform our analysis for the DUNE atmospheric data, which is a conservative choice.
In particular, the {\it downward-going} $\nu_\tau$ events, together with the help of exact identification and reconstruction of tau leptons, can be highly efficient in proging $g_{\alpha \tau}$ couplings.
We observed that including the NC background does not change our conclusions
significantly. Additionally, FLArE100 and FASER$\nu$2 have the potential to significantly enhance the current bounds on $g_{e\tau}$ and $g_{\mu\tau}$, while also slightly improving the constraints on $g_{ee}$ and $g_{e \mu}$. Notably, in the case of $g_{ee}$ and $g_{e\mu}$, above a few hundred MeV, SHiP is more sensitive in probing the couplings due to the larger number of produced $D_s$ mesons 
compared to the atmospheric case and the background contamination by conventional $\nu_\mu$ and $\nu_e$ in the atmosphere.

It is worth noting that in our analysis of DUNE atmospheric neutrinos we have assumed a fixed flux shape.
Inclusion of flux shape uncertainty could significantly modify the obtained sensitivities on $g_{ee}$ and $g_{e \mu}$. On the other hand, the sensitivities on $g_{e\tau}$ and $g_{\mu\tau}$ will be highly reliable on the direction of the tau neutrino and the shape of the background instead, so our results are quite robust on the flux uncertainties.
Our analysis results here can guide future experimental searches for new physics beyond the Standard Model. It is also important to note that the sensitivities
obtained in this study are based on the 
reference scenario with a sub-GeV level new gauge boson.
Other theoretically well-motivated models that predict different types of 
secret interactions  may result in different  
sensitivities as well as the astrophysical and cosmological constraints, which is worth to be studied in a future work.
In conclusion, the currently on-going and future tau neutrino experiments
such as DUNE, FLArE100, FASER$\nu$2, SND$@$LHC, and SND$@$SHiP, have great potential to search for a hidden interaction between neutrinos mediated by a new light sub-GeV gauge boson. 
In this regard, we emphasize on the importance of increasing the efficiency and accuracy of tau neutrino reconstruction in
searching for BSM interacting with tau neutrinos and encourage the experimental colleagues in this direction.


\subsection*{Acknowledgments}
Authors are grateful to Osamu Seto for useful remarks. 
Hospitality at APCTP during the program “Dark Matter as a Portal to New Physics” is kindly acknowledged.
This work is supported by the National Research Foundation of Korea (NRF-2020R1I1A3072747 and NRF-2022R1A4A5030362).

 \bibliography{ref}

\end{document}